# Hybrid 1D Plasmonic/Photonic Crystals are Responsive to *Escherichia Coli*

Giuseppe Maria Paternò[1†*], Liliana Moscardi[1,2†], Stefano Donini[1], Davide Ariodanti[3], Ilka Kriegel[4], Maurizio Zani[2], Emilio Parisini[1], Francesco Scotognella[1,2] and Guglielmo Lanzani[1,2*]

[1]Center for Nano Science and Technology@PoliMi, Istituto Italiano di Tecnologia, Via Giovanni Pascoli, 70/3, 20133 Milano, Italy;

[2]Dipartimento di Fisica, Politecnico di Milano, Piazza Leonardo da Vinci 32, 20133 Milano, Italy.

[3]Dipartimento di Chimica, Materiali e Ingegneria Chimica "Giulio Natta", Piazza Leonardo da Vinci 32, 20133 Milano, Italy.

[4]Department of Nanochemistry, Istituto Italiano di Tecnologia (IIT), via Morego, 30, 16163 Genova, Italy

[†]These authors contributed equally to this work

[*] Corresponding authors

## Abstract

Photonic crystal-based biosensors hold great promise as valid and low-cost devices for real-time monitoring of a variety of biotargets. Given the high processability and easiness of read-out even for unskilled operators, these systems can be highly appealing for the detection of bacterial contaminants in food and water. Here, we propose a novel hybrid plasmonic/photonic device that is responsive to *Escherichia coli*, which is one of the most hazardous pathogenic bacterium. Our system consists of a thin layer of silver, a metal that exhibits both a plasmonic behavior and a well-known biocidal activity, on top of a solution processed 1D photonic crystal. We attribute the bio-responsivity to the modification of the dielectric properties of the silver film upon bacterial contamination, an effect that likely stems from the formation of polarization charges at the Ag/bacterium interface within a sort of "bio-doping" mechanism. Interestingly, this triggers a blue-shift in the photonic response. This work demonstrates that our hybrid plasmonic/photonic device can be a low-cost and portable platform for the detection of common contaminants in food and water.

# Introduction

The integration of sensing elements with photonic crystals (PhCs) allows a simple readout of the detection event, often based on color changing, fostering applications in portable, and cheap technologies (*1–5*). For instance, photonic sensing might be of particular interest for the detection of contaminants or pathogenic bacteria in food and water, as in this case the vast majority of the existing detection systems are relatively time- and money-consuming mostly due to complexity of the read-out (*6, 7*).

Briefly, the periodicity in the dielectric constant along 1, 2 or 3 spatial dimensions gives rise to a forbidden gap for photons (stop-band) of specific wavelengths that, in turns, confers structural reflection colors to the material (*8*). Focusing on the simplest case of one-dimensional photonic crystals (also known as Bragg stacks, BSs) in which the stop-band arises from the alternation of layers with high/low refractive index, the structural color can be easily tuned by varying either the dielectric contrast or the periodicity of the alternated layers (or both). This can be achieved by the introduction of a medium within the 1D structure, as enabled by porosity at the meso/nanoscale in the BSs (*9–16*). To further enhance selectivity and to detect large or complex analytes (i.e. bacteria and biomolecules) it is also possible to chemically functionalize the surface of porous BS (*17–19*), although such a step would hamper easy scalability of the process. To this end, the fabrication of photonic sensors from scalable and low-cost procedures allowing fast and reliable detection of contaminants (i.e. in food and water) is highly desirable (*20*).

In this context, our recent work has been focused on the development of responsive BSs made of alternating layer of dielectric materials and electro-optical responsive plasmonic materials, which are fabricated from easy and low-cost solution-based processes. In particular, the integration of metal plasmonic systems in PhCs provides both high sensitivity to environmental changes as well peculiar sensing capabilities (*21*, *22*) due to the specific metal interactions (*23–25*). Carrier density modulation results in the change of the refractive index that ultimately determines the photonic stop-band, thus offering a handle for easy optical detection. We have shown that such peculiar feature of plasmonic materials can be exploited to build-up electro-optical switches based on the photonic reflection shift upon photo-electro doping of indium tin oxide (ITO) nanoparticles (NPs) in $SiO_2$/ITO and $TiO_2$/ITO photonic crystals (*26*, *27*), and electro doping of silver NPs in $TiO_2$/Ag crystals (*28*). Furthermore, the specific interactions occurring at the metal surface in contact with the analytes can be exploited for label-free and low-cost (bio)sensing purposes (*29*, *30*). It is well-known that silver films and NPs exhibit antibacterial properties (*31–35*). Although the exact antibacterial mechanism is still under debate (*36*) (see supplementary information section for a brief discussion on Ag bactericidal mechanism), many reports agree that electrostatic attraction is crucial for the Ag adhesion to the bacterial membrane and to the consequent bactericidal activity possibly mediated by transmembrane ion penetration (*35*, *37–39*). Interestingly, this might lead to a modification of Ag charge carriers density and plasmon resonance upon bacteria/Ag interaction, for example as a result of polarization (*40*) charges accumulating at the bacteria/Ag interface.

Here, we show that a novel hybrid plasmonic/photonic device consisting of a thin layer of silver deposited on top of a solution processed BS is responsive to one of the most common bacterial contaminant, namely *Escherichia coli*. Our data suggest that the increase in the plasmon charge density likely originates from the formation of polarization charges at the bacterium/Ag interface, resulting in a blue-shift of the plasmon resonance. This eventually determines a change in the photonic read-out (blue-shift) that translates the plasmonic effect occurring in the UV/blue (330 - 440 nm) to the more convenient spectral region (600-530 nm). These promising results indicate that hybrid plasmonic/photonic PhCs can represent a novel class of low-cost devices responsive to common contaminants in food and water.

## Results and discussion

### Hybrid plasmonic/photonic devices

The multilayered 1D photonic structures show the expected structural color in reflection (5 × $SiO_2$/$TiO_2$ bilayers) as shown in figure 1a-b, while electron microscopy images are reported in figure S1a. The BSs were fabricated via simple spin-coating deposition of the respective aqueous colloidal dispersions. This is a key point in the view to scale the process by means of large-area and low-cost deposition techniques, such as ink-jet printing and roll-to-roll. On top of the dielectric BS we deposited a thin layer of silver (8 nm, Fig. S1b for the electron microscopy image), to exploit both the plasmonic behaviour and the marked and well-documented bioactivity. The thin silver layer is in-fact a defective cap layer of the photonic crystal that affects the optical response of the BS through the silver free carrier density (Drude model) (*41*). Therefore, the main idea here is to exploit the possible change in the silver complex dielectric function driven by Ag/bacteria interaction to modify the dielectric properties at the BS/metal interface and, thus, the BS optical read-out. To this end, we firstly selected the minimum Ag thickness achievable with our deposition apparatus to localize strongly the plasmonic response in close proximity to the BS interface. To observe both the plasmonic and the photonic contributions to the overall sample transmission and disentangle them, we carried out measurements as a function of incidence angle (Fig. 1c). These data show a blue-shift of the photonic band-gap (PBG, 586 nm at 0°) by increasing the angle in agreement with the Bragg-Snell law (inset Fig. 1c) (*21*), while the plasmonic peak at 500 nm does not display any angular dependence.

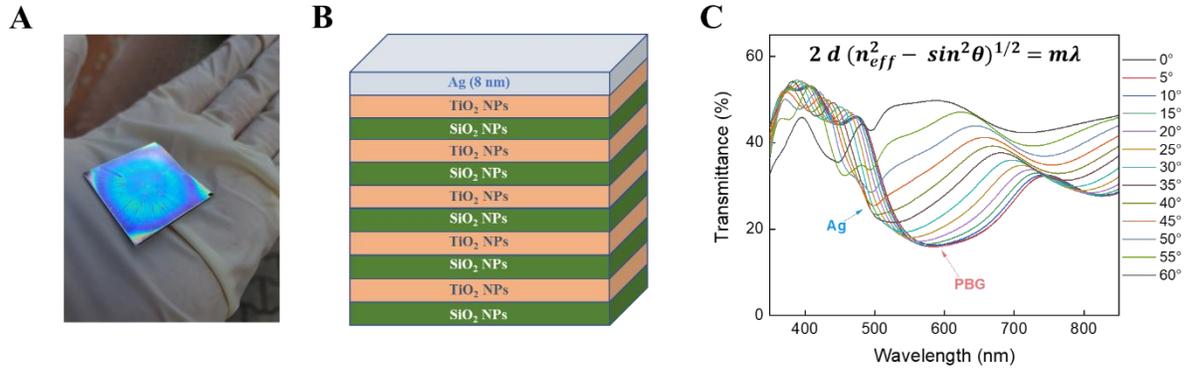

**Figure 1. Hybrid plasmonic/photonic devices.** (**A**) Picture of the fabricated 1D photonic crystal with a 8 nm silver capping layer and (**B**) sketch of the multilayered structure. (**C**) Light transmission of the Ag/(SiO$_2$/TiO$_2$)$_5$ photonic crystals as a function of the light incidence angle (inset Bragg-Snell equation). PBG stands for photonic band-gap

## Plasmonic response upon *E. coli* contamination

To evaluate the effect of bacteria on the optical properties of silver, we first exposed Ag films to LB only (control experiment) and then to *E. coli* (Fig. 2a) in agar plate, as described in the experimental section. We observe that the sample exposed to LB undergoes a substantial red-shift (+ 60 nm) that is likely due to the infiltration of the aqueous culture medium across the silver grains, leading to an increase in the effective refractive index (*42*). On the other hand, when the Ag layer is contaminated with *E. coli* in LB medium we note the concomitant increase (+25%) of the plasmonic absorption at the high energy side (330 - 440 nm) and an attenuated red-shift (+ 35 nm) with respect to LB only exposure. Taken together, these data indicate an overall blue-shift (- 25 nm) in the Ag plasmonic response upon contamination with *E. coli*. Furthermore, to mimic exposure to contaminated liquid samples, we dipped them either in LB medium (control) or in an LB/*E. coli* mixture with increasing bacterial loading (0.1, 0.5 and 1.2 OD$_{600nm}$) and measured their plasmonic response (Fig. S2). Here, we essentially observed an analogous effect, with an increased plasmonic blue-shift upon exposure to bacteria that, interestingly, can be already noticed at the lowest loading (-15 nm at 0.1 OD$_{600nm}$). In this scenario, we hypothesize that the bacteria-induced blue-shift in the plasmon resonance could stem from the formation of polarization charges at the silver-bacterium interface, i.e. negative on bacterial membrane (*38*) and positive charges on Ag surface respectively (see Fig. 2c), finally leading to an overall increase of the charge carrier density (*26*, *28*). By employing the Lorentz-Drude model (Fig. 2d), we estimated that the 25 nm blue-shift observed experimentally corresponds to a 15% increase in the charge carried density of Ag in LB medium.

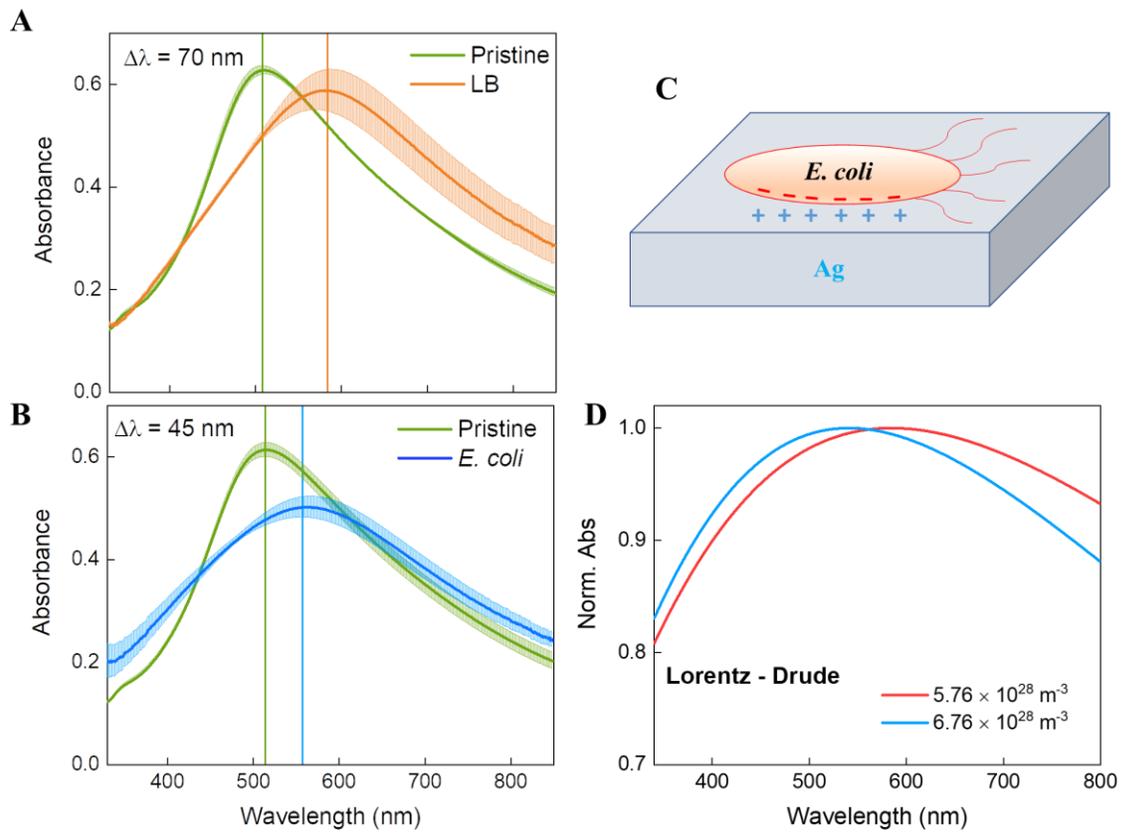

**Figure 2. Plasmonic response upon bacterial contamination.** (**A**) Average optical absorption (with standard deviation) of Ag thin films (8 nm on glass) before (green) and after exposure to either LB (orange line) or to (**B**) LB/*E. coli* (blue line) on agar plate (12 samples, two replicas). (**C**) Pictorial representation of the hypothesized mechanism giving rise to the plasmonic blue-shift (polarization accumulation charge). (**D**) Calculated transmission of the Ag layer for the pristine and "bio-doped" Ag layer in LB medium obtained by using the Lorentz - Drude model.

## Photonic response upon *E. coli* contamination

After having investigated the Ag plasmonic response of metallic film alone upon contamination with *E. coli*, we proceeded to study how this is affecting the all optical read-out in presence of the photonic crystal (Fig. 3). Qualitatively, both exposure to LB or *E. coli* + LB leads to the same behavior, namely decreased transmittance, resonance broadening (20 nm) and red-shift of the stop-band. Again we have two physical phenomena concurring to these changes: i) infiltration of LB inside the porous BS architecture, leading to an enhanced effective refractive index and ii) the modulation of the plasmon resonance in the top metal layer. The magnitude of the PBG red-shift, however, results decreased in presence of *E. coli* in LB, featuring an average shift of 5 nm compared to LB alone with + 15 nm. Interestingly, this yields an overall PBG blue-shift of -10 nm upon contamination, which essentially translates the plasmonic effect observed in the UV/blue region into the green/red part of the spectrum. The presence of a well-defined zone of inhibition matching the shape of our samples in conformal contact with the agar medium inoculated with *E. coli* confirms the anti-bacterial activity of our Ag layer (Fig. S3). Furthermore, this effect seems to be confined within the sample area, suggesting that silver does not appreciably detach from the surface and diffuse in all the agar plate as it occurs with silver NPs (*43*), which may be of importance when considering possible applications of these devices in food packaging (*44*). To study the optical response of Ag/PhCs in contact with contaminated liquid specimens, we also immersed our samples in LB medium or *E. coli*/LB (10 minutes) and measured their optical transmission (Fig. S4). Although in this case the effect might appear less evident due the massive and unavoidable LB infiltration throughout the porous structure, we can still observe a decreased red-shift for the contaminated samples when compared to the LB case, an effect that can be already noted at the lowest bacterial loading (-10 nm for 0.1 $OD_{600nm}$) in analogy with the plasmonic response. Finally, as a control experiment, we repeated the same procedure in contaminated agar plate on identical 1D photonic crystals but without the top Ag layer, in order to further prove the role of the plasmonic material in the bio-responsivity of our device (Fig. 3c). Here, we could not discriminate any difference between the samples exposed to LB or to LB/*E. coli*, with the exception of the usual read-shift and broadening of the stop-band that are connected to liquid condensation and infiltration throughout the porous structure. This confirms that the upper plasmonic layer represents the responsivity element of our hybrid plasmonic/photonic device, as the modification of the dielectric properties at the Ag/BS interface, which in turns is brought about by Ag/bacteria interaction, and governs the total optical read-out.

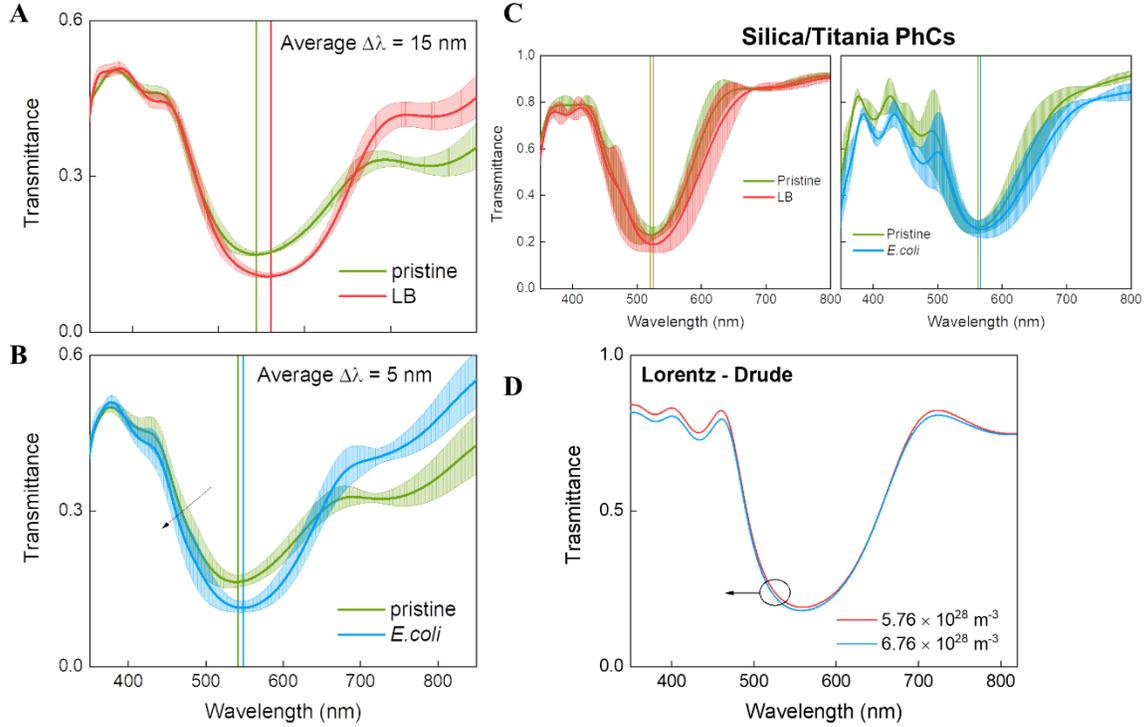

**Figure 3. Hybrid plasmonic/photonic response of hybrid Ag (8 nm) /1D PhCs upon contamination with *E. coli*.** (**A**) Average transmittance spectrum (with standard deviation) of Ag/1D PhC after exposure to the LB medium (red-line) and (**B**) *E. coli* (blue-line). (**C**) Average shift at the stop-band maximum for the LB and *E. coli* contaminated PhCs. Data were averaged over two sets of measurement (six samples per measurement). (**C**) Average transmittance spectrum (with standard deviation) of Silica/Titania 1D PhC after exposure to the LB medium (red-line) and E. coli (blue-line). (**D**) Calculated transmission spectrum for the Ag/BS structure in LB medium for $n = 5.76 \times 10^{28}\ m^3$ and $6.76 \times 10^{28}\ m^3$.

To obtain insights into the mechanism underpinning the photonic shift, we calculated the transmission spectra as a function of the silver carrier density (Fig. 3d). For this, we combined the transfer matrix method to model the alternating refractive index of the periodic structure, with the Maxwell-Garnett effective medium approximation for the description of the effective refractive indexes of the $SiO_2/TiO_2$ layers soaked with LB (see experimental section), In addition, we made use of the Lorentz-Drude model to account for the plasmonic contribution to the overall dielectric response of the device. We then proceeded to the simulation of the transmission spectra for $n = 5.76 \times 10^{28} \, m^3$ and $6.76 \times 10^{28} \, m^3$, with the former charge carrier density accounting for pristine Ag and the latter for the "bio-doped" film in the LB infiltrated photonic structure. Indeed, while an increased carrier density induces a large blue-shift of the plasmon resonance (25 nm) as it has been shown in the previous section, the PBG exhibits a less obvious 5 nm blue-shift. Notably, such a behavior corroborates our experimental data, at least from the qualitative point of view. Such discrepancy can be probably attributed to the contribution of a different effect that intensifies the dielectric mechanism, for instance a strong field enhancement and confinement at the metal/PhC interface (i.e. Tamm optical modes) (*45*, *46*) that cannot be taken into account by our model. To preliminary assess this possible scenario, we also studied the photonic response of our hybrid PhCs with a top Ag layer of 16 nm (Fig. 4). In this case, we observe a more pronounced contribution of the Ag plasmon to the overall transmission of the hybrid structure due to its higher optical density than in the previous samples (Fig. 4a). Remarkably, despite the plasmon resonance still shows a total 10 nm blue-shift after contamination when compared with the LB-exposed samples, the PBG shift is limited to - 5 nm. This might imply the involvement of an enhancement mechanism when Ag thickness is kept at relatively low value, likely due to the strong confinement of the plasmon at the dielectric interface (*45–47*). Further experiments are needed to elucidate such a mechanism.

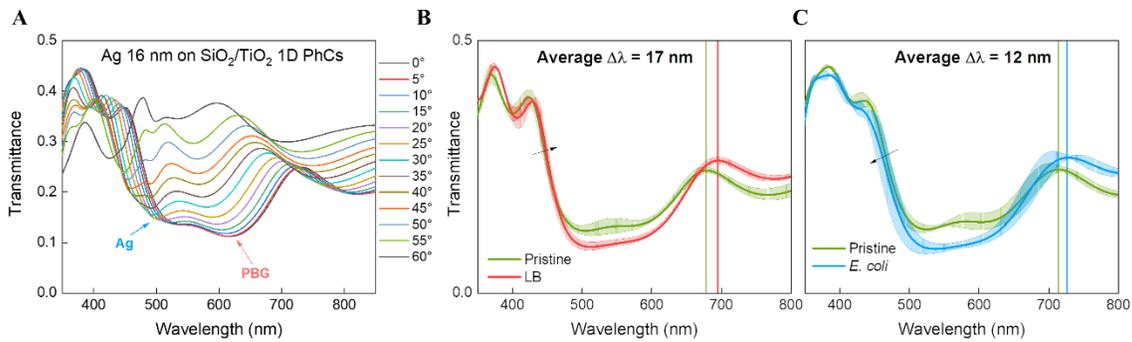

**Figure 4. Plasmonic/photonic response of hybrid Ag (16 nm)/1D PhCs.** (**A**) Light transmission of the Ag/(SiO$_2$/TiO$_2$)$_5$ photonic crystals as a function of the light incidence angle. (**B**) Average transmittance spectrum (with standard deviation) of Ag/1D PhC after exposure to the LB medium (red-line) and (**B**) E. coli (blue-line). Data were averaged over two sets of measurement (six samples per measurement).

## Conclusions

To summarize, we have shown that a novel hybrid plasmonic/photonic device consisting of a thin layer of the plasmonic and biocidal silver on top of a 1D photonic crystal can be responsive against one of the most hazardous bacterial contaminant in food and water, *Escherichia coli*. Our data points towards a scenario in which the polarization charge at the Ag/bacterium interface causes an increase of the metal charge carrier density that leads to a plasmon blue-shift, within a sort of "bio-doping" mechanism. This, in turns, leads a photonic blue-shift in the visible spectral range. It is worth adding that the photonic band-gap can be placed in any spectral region by a judicious choice of the layer thicknesses, allowing one to translate the photonic read-out in the most convenient part of the spectrum. This, taken together with the quick processability from solution and easiness of the read-out, makes these devices promising for low-cost and real-time monitoring of contaminants in food and water.

## Experimental section

**Photonic crystals fabrication.** Porous 1D PhCs were fabricated by alternating layers of $SiO_2$ and $TiO_2$ via spin-casting deposition from their colloidal aqueous dispersions following the procedure employed in past experiments.(*26*, *27*) Firstly, we suspended the $TiO_2$ (Gentech Nanomaterials, average size 5 nm) and $SiO_2$ nanoparticles (Sigma Aldrich LUDOX SM-30, average size 8 nm) in MilliQ distilled water to obtain a concentration of 5 wt. %. The dispersions were then sonicated for 2 h at 45 °C (Bandelin SONOREX Digital 10 P) and filtered with a 0.45 μm PVDF filter. The glass substrates were previously cleaned by means of ultra-sonication in isopropanol (10 min) and acetone (10 min), and then subjected to an oxygen plasma treatment (Colibrì Gambetti, 10 min) to increase wettability. During the fabrication of the PhCs, the dispersions were continuously kept in sonication at 45 °C to maintain homogeneity of the dispersions during the whole process. The fabrication of the crystals was performed by alternating the deposition of the two materials through spin-coating (Laurell WS-400-6NNP-Lite) with a speed of 2000 rpm. After each deposition, the samples were annealed on a hot-plate for 20 min at 350 °C. Finally, we deposited an 8 nm-thick silver layer on top of the photonic structure (or glass substrate only) via thermal evaporation (MBRAUN metal evaporator).

**Bacterial culture.** A single colony from the *Escherichia coli Rosetta* (DE3) strain carrying a pET23a (+) plasmid (Novagen) was inoculated in Luria Bertani (LB) broth in the presence of Ampicillin (50 μg/ml) and incubated overnight at 37ºC with shaking at 200 rpm until stationary phase was reached. Then, bacterial suspension turbidity (expressed as optical density at 600nm; O.D. 600) was diluted to O.D. 600 ~ 0.5 in LB broth (no antibiotic). The suspension (500 μL) was spread over an LB agar plate. The Ag/PhCs were placed at the center of the Petri dish with the top silver layer facing the contaminated surface (or LB only for the control experiment) and incubated for 24 h at 37 °C. We also dipped the devices in either LB only (control) or an LB/*E.coli* mixture to mimic exposure to contaminated liquid samples (0.1, 0.5 and 1.2 O.D.). The same protocol was also repeated for silver thin films on glass substrates in order to understand the effect of bacteria exposure on the silver plasmon resonance. Data were averaged over two sets of measurements (six samples per measurement).

**Optical characterization**. The optical characterization was performed using a spectrophotometer (Perkin Elmer Lambda 1050 WB), measuring the percentage loss of transmittance after exposure to LB or LB/*E. coli*. To disentangle the photonic from the plasmonic contribution to the overall transmission of the sample, we recorded the transmission as a function of the incidence angle, showing a blue-shift of the photonic stop-band upon increase of the angle, in accordance with the Bragg-Snell law.

**Scanning electron microscopy**. We used a Tescan MIRA3. The measurements were performed at a voltage of 5 kV and backscattered electrons were detected. The sample was covered with carbon paste to improve conductivity

**Transfer matrix method.** *Refractive indexes of the Ag layers, and the SiO$_2$ and TiO$_2$ layers composing the photonic crystal:* we employ the combination of Drude model and Lorentz model to describe the plasmonic response(*48–50*). Thus, the frequency dependent complex dielectric function of silver can be written as:

$$\varepsilon_{Ag,\omega} = \varepsilon_\infty - \frac{\omega_D^2}{\omega(\omega+i\Gamma_D)} - \frac{\Delta\varepsilon\,\omega_L^2}{\omega^2-\omega_L^2+i\Gamma_L\omega} \qquad (1)$$

where $\omega$ is in *rad/s*. The first term $\varepsilon_\infty$ is the high frequency dielectric constant ($\varepsilon_\infty = 0.1148$ in this work(*48*)). The second term is the Drude model part, where $\omega_D$ is the Drude plasma frequency:

$$\omega_D = \sqrt{\frac{Ne^2}{m^*\varepsilon_0}} \tag{2}$$

with $N$ number of charges, $e$ the electron charge, $m^*$ the effective mass and $\varepsilon_0$ the vacuum dielectric constant. For silver, we use $N = 5.76 \times 10^{28}\ charges/m^3$ and $m^* = 0.96/m_0\ kg$ (*51*). $\Gamma_D$ is the damping coefficient (in this work $\Gamma_D = 7.055 \times 10^{15}\ rad/s$ (*48*)).

The third term of Equation 1 is the Lorentz model part with a single Lorentz term. $\Delta\varepsilon$ is the oscillator strength, $\omega_L$ the Lorentz plasma frequency, $\Gamma_L$ is the damping coefficient (in this work $\Delta\varepsilon = 3.6276$, $\omega_L = 1.5812 \times 10^{16}\ rad/s$, $\Gamma_L = 1.0463 \times 10^{14}\ rad/s$ (*48*)).

The wavelength dependent refractive index of TiO$_2$ can be written(*52*):

$$n_{TiO_2}(\lambda) = \left(4.99 + \frac{1}{96.6\lambda^{1.1}} + \frac{1}{4.60\lambda^{1.95}}\right)^{1/2} \tag{3}$$

where $\lambda$ is the wavelength [in micrometers, and $\varepsilon_{TiO_2}(\lambda) = n_{TiO_2}^2(\lambda)$]. Instead, the wavelength dependent refractive index of SiO$_2$ can be described by the following Sellmeier equation (*53*):

$$n_{SiO_2}^2(\lambda) - 1 = \frac{0.6961663\lambda^2}{\lambda^2-0.0684043^2} + \frac{0.4079426\lambda^2}{\lambda^2-0.1162414^2} + \frac{0.8974794\lambda^2}{\lambda^2-9.896161^2} \quad (4)$$

where $\lambda$ is the wavelength in micrometers [also for SiO$_2$ $\varepsilon_{SiO_2}(\lambda) = n_{SiO_2}^2(\lambda)$].

Taking into account the infiltration of LB in the silver layer and in the silica and titania layers of the photonic crystal, we determine the effective dielectric function of the SiO$_2$:air layer (we call it $\varepsilon_{eff2,\omega}$) by using the Maxwell Garnett effective medium approximation (*54, 55*):

$$\varepsilon_{eff,\omega} = \varepsilon_{LB} \frac{2(1-f)\varepsilon_{LB}+(1+2f)\varepsilon_\omega}{2(2+f)\varepsilon_{LB}+(1-f)\varepsilon_\omega} \quad (5)$$

where $f$ is the filling factor of the silver, SiO$_2$, or TiO$_2$ layers and $\varepsilon_\omega$ is the dielectric function of the silver, SiO$_2$, or TiO$_2$ layers. In this study we choose $f_{Ag} = 0.5$ ; $f_{SiO_2} = 0.6$ ; $f_{TiO_2} = 0.6$. The dielectric constant in the visible range for LB is approximated to the one of water, thus $\varepsilon_{LB} = 1.769$.

*Transmission of the multilayer photonic crystal*: We use the two effective refractive indexes of the Ag:LB, SiO$_2$:LB, TiO$_2$:LB layers to study the light transmission through the photonic structure by employ the transfer matrix method (*56, 57*). For a transverse electric (TE) wave the transfer matrix for the $k$th layer is given by

$$M_k = \begin{bmatrix} \cos\left(\frac{2\pi}{\lambda}n_k d_k\right) & -\frac{i}{n_k}\sin\left(\frac{2\pi}{\lambda}n_k d_k\right) \\ -in_k \sin\left(\frac{2\pi}{\lambda}n_k d_k\right) & \cos\left(\frac{2\pi}{\lambda}n_k d_k\right) \end{bmatrix} \quad (6)$$

with $n_k$ the refractive index and $d_k$ the thickness of the layer. In this study the thickness of the Ag:LB layers is 8 nm, while the thickness of the SiO$_2$:LB layers and TiO$_2$:LB layers is 100 nm.

The product $M = M_1 \cdot M_2 \cdot \ldots \cdot M_k \cdot \ldots \cdot M_s = \begin{bmatrix} m_{11} & m_{12} \\ m_{21} & m_{22} \end{bmatrix}$ gives the matrix of the multilayer (of $s$ layers). The transmission coefficient is

$$t = \frac{2n_s}{(m_{11}+m_{12}n_0)n_s+(m_{21}+m_{22}n_0)} \quad (7)$$

with $n_s$ the refractive index of the substrate (in this study $n_s = 1.46$) and $n_0$ the refractive index of air. Thus, the light transmission of the multilayer photonic crystal is

$$T = \frac{n_0}{n_s}|t|^2 \quad (8)$$

## Author information

**Corresponding authors**

E-mail: Giuseppe.Paterno@iit.it ; Guglielmo.Lanzani@iit.it

**Authors contribution**

G.M.P., L.M., S.D. and D.A. carried out the experiments. G.M.P. wrote the manuscript, together with L.M., S.D., G.L. and F.S. I.K., M.Z. and F.S. run the simulations. E.M supervised the microbiology work. G.M.P., G.L. and F.S. conceived and supervised the work. G.M.P and L.M. contributed equally to this work.

**Acknowledgements**

This project has received funding from the European Research Council (ERC) under the European Union's Horizon 2020 research and innovation programme (grant agreement No. [816313]). We thank Dr. Stefano Perissinotto for the scanning electron microscopy experiments on 1D PhCs cross section and Ag films.

**Conflict of interest**

The authors declare no conflict of interest.

**Keywords**

Responsive Photonics; Plasmonics; Bacterial contaminants; Antibacterial; Silver

*Supplementary information for*

# A Hybrid 1D Plasmonic/Photonic Crystals are Responsive to *Escherichia Coli*

Giuseppe Maria Paternò[1†*], Liliana Moscardi[1,2†], Stefano Donini[1], Davide Ariodanti[3], Ilka Kriegel[4], Maurizio Zani[2], Emilio Parisini[1, *] Francesco Scotognella[1,2] and Guglielmo Lanzani[1,2*]

[1]Center for Nano Science and Technology@PoliMi, Istituto Italiano di Tecnologia, Via Giovanni Pascoli, 70/3, 20133 Milano, Italy;

[2]Dipartimento di Fisica, Politecnico di Milano, Piazza Leonardo da Vinci 32, 20133 Milano, Italy.

[3]Dipartimento di Chimica, Materiali e Ingegneria Chimica "Giulio Natta", Piazza Leonardo da Vinci 32, 20133 Milano, Italy.

[4]Department of Nanochemistry, Istituto Italiano di Tecnologia (IIT), via Morego, 30, 16163 Genova, Italy

[†]These authors contributed equally to this work

[*]Corresponding authors

**This file includes**: Discussion on the biocidal activity of silver; scanning electron microscopy images of the silver film and cross-section of the 1D photonic crystal; UV-Vis absorption spectra of the silver films after dipping in liquid LB and *E. coli*/LB mixture; zone of inhibition in the agar plate inoculated with *E. coli*; UV-Vis transmission spectra of the Ag/BSs samples after dipping in liquid LB and *E. coli*/LB mixture.

# Biocidal activity of silver

The antimicrobial properties of silver are well-known, and are exploited in a variety of everyday life applications(*1*, *2*). For this reason, silver thin film and nanoparticles (NPs) have been employed in a range of applications as antibacterial agent, which can be also useful for addressing the problem of antibiotic resistance that is occurring in the last decade (*3*). However, a unified and definitive explanation to describe this property has not been provided yet (*4*).

Various studies have related the cytotoxic effect of Ag to a combination of different reactions that take place within the prokaryotic cell(*4*). Furthermore, it has been observed that the anti-bacterial efficacy depends on the type of bacteria(*5*). The bacteria can be cataloged in Gram-positive and Gram-negative, based on the conformation of their membrane. In the former, the cell wall is composed of a thick layer of peptidoglycans, while the Gram-negatives have an outer membrane formed mainly by lipopolysaccharides and an inner one of peptidoglycans, much thinner than the previous ones(*6*). Given the larger size of the Gram-positive cell wall, the endocytosis of the nanoparticles is more difficult, and therefore less effective for this kind of bacteria(*3*).

The cytotoxic effect of silver nanoparticles is more or less marked depending on concentration, shape, time and size(*3*, *7*, *8*). NPs with dimensions smaller than 10 nm are more easily absorbed by endocytosis from the cell and interact with lysosomes and endosomes, in fact, thanks to their large specific surface area they are particularly reactive(*3*, *7*, *9*). The acidic environment present inside the lysosomes favors chemical reactions that increase the presence of ROS (reactive oxygen species) and of superoxide anion ($O^{2-}$) produced by them, leading to an arise of oxidative stress. ROS cause an imbalance between the cell's ability to eliminate reactive intermediates and oxygen production(*9*). Hydrogen peroxide ($H_2O_2$) contained in the ROS reacts with Ag NPs leading to the formation of Ag ions: $2Ag + H_2O_2 + 2H^+ \rightarrow 2Ag^+ + 2H_2O$ (*7*). Moreover, $H_2O_2$ can lead to the formation of ·OH, considered one of the most oxidative ROS, able to oxidize the whole cell. Nanoparticles and silver ions can escape from lysosomes, increasing intracellular ROS concentration(*10*). Ag NPs and Ag ions reduce glutathione, thioredoxin, superoxide dismutase and thioredoxin peroxidase, as they react with thiol groups, contained in most of the cell including cytoplasm, mitochondria and cell membrane. Damage to the cell wall causes an increase in its permeability and a cytoplasm leak, thus leading to necrosis(*11*). Also the lysosomal membrane rupture causes a lysosome-mediated apoptosis, pouring the cathepsins into the cytoplasm(*11*). Ag NPs and $Ag^+$ cause damage to mitochondria by inhibiting the production of adenosine triphosphate (ATP), increasing oxidative stress, interfering with mitochondria impairs electron transfer, causing swelling and acceleration of mitochondrial respiration itself and leading to apoptosis(*11*, *12*). Nuclear pore complex has an average diameter of 9-10 nm, hence small AgNPs can penetrate and deposit inside with a subsequent production of ROS that damages the DNA and generates chromosomal abnormalities(*11*). Also $Ag^+$ has been seen to cause problems to DNA and induce apoptosis of the cells(*7*, *10*).

(a) 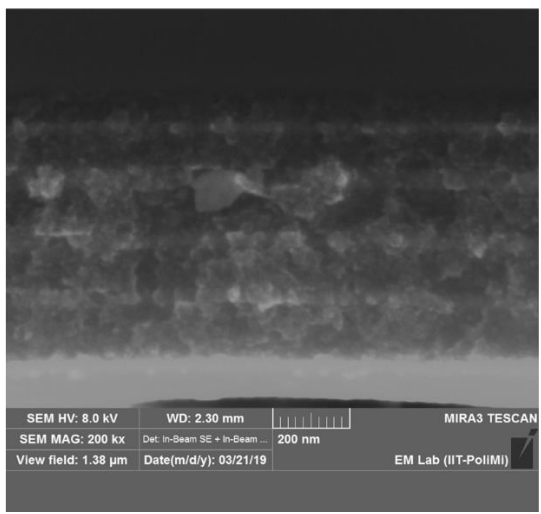 (b) 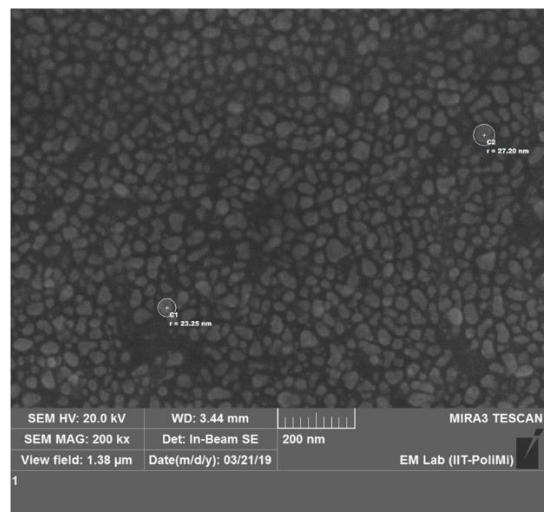

**Figure S1** (**A**) Scanning electron microscopy (SEM) image of Ag/BS sample (cross section). (**B**) SEM of the Ag layer exhibiting the typical granular morphology.

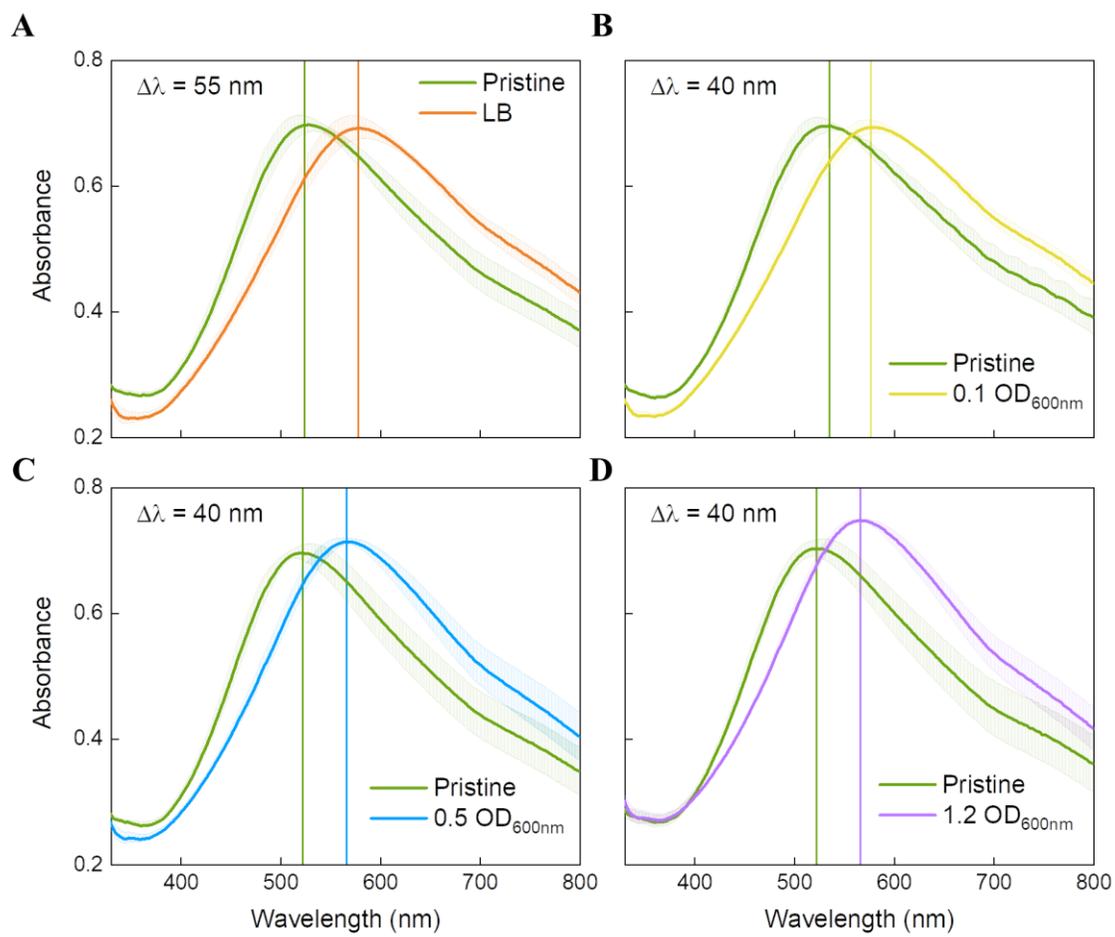

**Figure S2**. Average optical absorption of a Ag thin film (8 nm on glass) before and after dipping in before and after dipping in (**A**) LB, (**B**) E. coli at 0.1 $OD_{600nm}$, (**C**) 0.5 $OD_{600nm}$, (**D**) 1.2 $OD_{600nm}$.

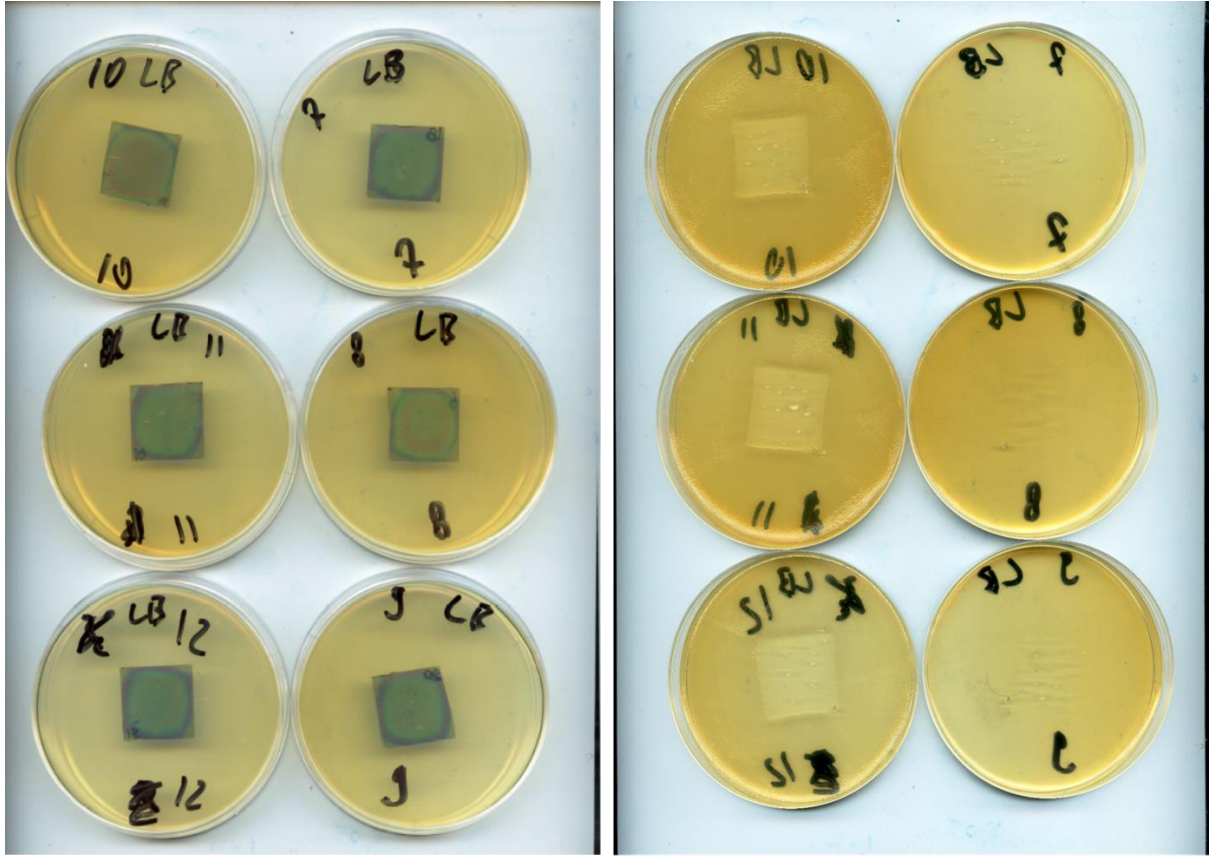

**Figure S3.** Picture of the Agar plate embedded with LB medium or inoculated with *E. coli* before (left) and after removal of our Ag/BSs.

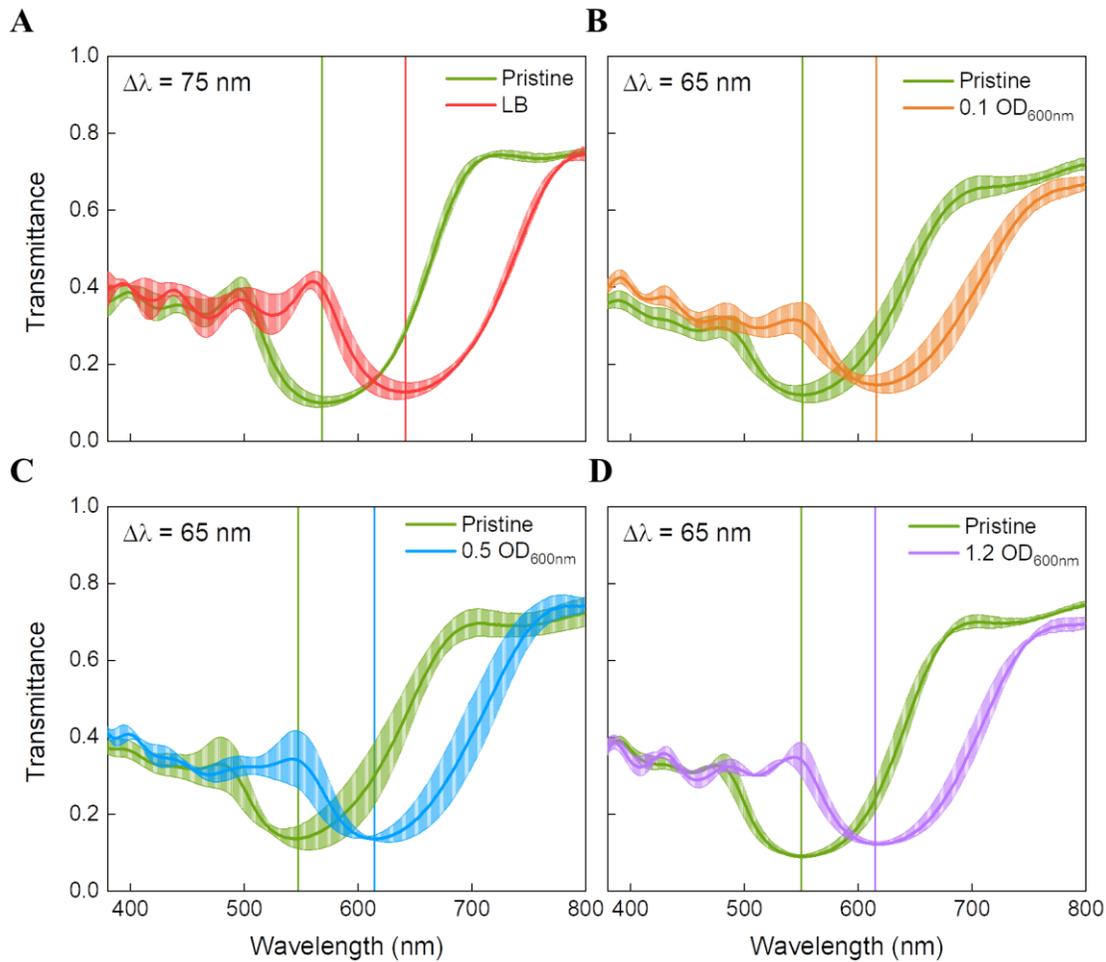

**Figure S4.** Average transmission spectrum of the Ag/BSs before and after dipping in (**A**) LB, (**B**) E. coli at 0.1 $OD_{600nm}$, (**C**) 0.5 $OD_{600nm}$, (**D**) 1.2 $OD_{600nm}$.